\newcommand{\CLASSINPUTtoptextmargin}{19mm}%
\newcommand{\CLASSINPUTbottomtextmargin}{43mm}%
\newcommand{\CLASSINPUTinnersidemargin}{12.9mm}%
\newcommand{\CLASSINPUToutersidemargin}{12.9mm}%
\documentclass[conference,10pt,a4paper]{IEEEtran}
\usepackage{amsmath}
\usepackage{cite}
\usepackage{times}
\usepackage{graphicx}
\usepackage{makecell}  
\usepackage{multirow}
\usepackage[none]{hyphenat}
\usepackage{float}
\usepackage{subfig}
\usepackage{iftex}
\usepackage{hyperref}  
\ifPDFTeX%
\usepackage{t1enc}
\usepackage{times}
\fi%
\ifXeTeX%
\usepackage{fontspec}
\fi%
\ifLuaTeX%
\usepackage{fontspec}
\fi%
\makeatletter

\def\@maketitle{\newpage
\bgroup\par\addvspace{0.5\baselineskip}\centering%
\ifCLASSOPTIONtechnote
   {\bfseries\large\@IEEEcompsoconly{\sffamily}\@title\par}\vskip 1.3em{\lineskip .5em\@IEEEcompsoconly{\sffamily}\@author
   \@IEEEspecialpapernotice\par{\@IEEEcompsoconly{\vskip 1.5em\relax
   \@IEEEtitleabstractindextextbox{\@IEEEtitleabstractindextext}\par
   \hfill\@IEEEcompsocdiamondline\hfill\hbox{}\par}}}\relax
\else
   \vskip0.2em{\EuMWtitlesize\ifCLASSOPTIONtransmag\bfseries\LARGE\fi\@IEEEcompsoconly{\sffamily}\@IEEEcompsocconfonly{\normalfont\normalsize\vskip 2\@IEEEnormalsizeunitybaselineskip
   \bfseries\Large}\@title\par}\vskip1.0em\par
   \ifCLASSOPTIONconference%
      {\@IEEEspecialpapernotice\mbox{}\vskip\@IEEEauthorblockconfadjspace%
       \mbox{}\hfill\begin{@IEEEauthorhalign}\@author\end{@IEEEauthorhalign}\hfill\mbox{}\par}\relax
   \else
      \ifCLASSOPTIONpeerreviewca
         {\@IEEEcompsoconly{\sffamily}\@IEEEspecialpapernotice\mbox{}\vskip\@IEEEauthorblockconfadjspace%
          \mbox{}\hfill\begin{@IEEEauthorhalign}\@author\end{@IEEEauthorhalign}\hfill\mbox{}\par
          {\@IEEEcompsoconly{\vskip 1.5em\relax
           \@IEEEtitleabstractindextextbox{\@IEEEtitleabstractindextext}\par\hfill
           \@IEEEcompsocdiamondline\hfill\hbox{}\par}}}\relax
      \else
         \ifCLASSOPTIONtransmag
           {\@IEEEspecialpapernotice\mbox{}\vskip\@IEEEauthorblockconfadjspace%
            \mbox{}\hfill\begin{@IEEEauthorhalign}\@author\end{@IEEEauthorhalign}\hfill\mbox{}\par
           {\vspace{0.5\baselineskip}\relax\@IEEEtitleabstractindextextbox{\@IEEEtitleabstractindextext}\vspace{-1\baselineskip}\par}}\relax
         \else
           {\lineskip.5em\@IEEEcompsoconly{\sffamily}\sublargesize\@author\@IEEEspecialpapernotice\par
           {\@IEEEcompsoconly{\vskip 1.5em\relax
            \@IEEEtitleabstractindextextbox{\@IEEEtitleabstractindextext}\par\hfill
            \@IEEEcompsocdiamondline\hfill\hbox{}\par}}}\relax
         \fi
      \fi
   \fi
\fi\par\addvspace{0.0\baselineskip}\egroup}

\def\EuMWtitlesize{\@setfontsize{\EuMWtitlesize}{24}{24pt}}
\def\EuMWauthorsize{\@setfontsize{\EuMWauthorsize}{11}{11pt}}
\def\EuMWaffilsize{\@setfontsize{\EuMWaffilsize}{10}{10pt}}
\def\EuMWcaptionsize{\@setfontsize{\EuMWcaptionsize}{9}{10pt}}
\def\EuMWbibsize{\@setfontsize{\EuMWbibsize}{8}{10pt}}

\def\@IEEEauthorblockNstyle{\EuMWauthorsize\@IEEEcompsocnotconfonly{\sffamily}\@IEEEcompsocconfonly{\large}}
\def\@IEEEauthorblockAstyle{\EuMWaffilsize\@IEEEcompsocnotconfonly{\sffamily}\@IEEEcompsocconfonly{\itshape}\@IEEEcompsocconfonly{\large}}
\def\@IEEEauthordefaulttextstyle{\EuMWauthorsize\@IEEEcompsocnotconfonly{\sffamily}\sublargesize}

\def\thebibliography#1{\section*{\refname}%
    \addcontentsline{toc}{section}{\refname}%
    \EuMWbibsize\@IEEEcompsocconfonly{\small}\vskip 0.3\baselineskip plus 0.1\baselineskip minus 0.1\baselineskip
    \list{\@biblabel{\@arabic\c@enumiv}}%
    {\settowidth\labelwidth{\@biblabel{#1}}%
    \leftmargin\labelwidth
    \advance\leftmargin\labelsep\relax
    \itemsep \IEEEbibitemsep\relax
    \usecounter{enumiv}%
    \let\p@enumiv\@empty
    \renewcommand\theenumiv{\@arabic\c@enumiv}}%
    \let\@IEEElatexbibitem\bibitem%
    \def\bibitem{\@IEEEbibitemprefix\@IEEElatexbibitem}%
\def\newblock{\hskip .11em plus .33em minus .07em}%
\ifCLASSOPTIONtechnote\sloppy\clubpenalty4000\widowpenalty4000\interlinepenalty100%
\else\sloppy\clubpenalty4000\widowpenalty4000\interlinepenalty500\fi%
    \sfcode`\.=1000\relax}

\long\def\@makecaption#1#2{%
\ifx\@captype\@IEEEtablestring%
\par\@IEEEtabletopskipstrut
\else
\@IEEEfigurecaptionsepspace
\fi
\setbox\@tempboxa\hbox{\normalfont\footnotesize {#1.}\nobreakspace\nobreakspace #2}%
\ifdim \wd\@tempboxa >\hsize%
\setbox\@tempboxa\hbox{\normalfont\footnotesize {#1.}\nobreakspace\nobreakspace}%
\parbox[t]{\hsize}{\normalfont\footnotesize\noindent\unhbox\@tempboxa#2}%
\else
\ifCLASSOPTIONconference \hbox to\hsize{\normalfont\footnotesize\hfil\box\@tempboxa\hfil}%
\else \hbox to\hsize{\normalfont\footnotesize\box\@tempboxa\hfil}%
\fi\fi
\ifx\@captype\@IEEEtablestring%
\@IEEEtablecaptionsepspace
\else
\fi}

\newlength\tablecaptiontotableskip
\newlength\figuretocaptionskip
\def\@IEEEfigurecaptionsepspace{\vskip\figuretocaptionskip\relax}%
\def\@IEEEtablecaptionsepspace{\vskip\tablecaptiontotableskip\relax}%

\def\abstract{\normalfont%
\@IEEEabskeysecsize\bfseries\textit{\abstractname}\,\bfseries\textit{---}\,%
\@IEEEgobbleleadPARNLSP}%

\def\IEEEkeywords{\normalfont%
\@IEEEabskeysecsize\bfseries\textit{\IEEEkeywordsname}\,\bfseries\textit{---}\,%
\@IEEEgobbleleadPARNLSP}%
\def\endIEEEkeywords{\relax\vspace{0.67ex}%
\par\if@twocolumn\else\endquotation\fi%
\normalsize\normalfont}%

\def\@IEEEauthorblockNtopspace{0ex}
\def\@IEEEauthorblockAtopspace{1mm}
\def\tablename{Table}
\def\thetable{\arabic{table}}
\def\IEEEkeywordsname{Keywords}
\def\subsubsection{\@startsection{subsubsection}{3}{\z@}{1.5ex plus 1.5ex minus 0.5ex}%
{0.7ex plus .5ex minus 0ex}{\normalfont\normalsize\itshape}}%
\newlength{\CPheadmatchindent}%
\def\@seccntformat#1{\hbox to\CPheadmatchindent{\csname the#1dis\endcsname}\hskip 0.1em \relax}
\IEEEilabelindentA \parindent
\IEEEilabelindent \IEEEilabelindentA
\IEEEelabelindent \parindent
\IEEEdlabelindent \parindent
\IEEElabelindent \parindent
\makeatother

\begin{document}
\raggedbottom
%
%
%
\title{RadarML: A Multi-Channel Over-the-Air Dataset of Common Radar Waveforms for ES Evaluation}
%
%
\author{%
\IEEEauthorblockN{%
Ryan White\IEEEauthorrefmark{1}, 
Adam Hosford\IEEEauthorrefmark{2},
Matthew A. Ritchie\IEEEauthorrefmark{1}
}
\IEEEauthorblockA{\IEEEauthorrefmark{1}Department of Electronic and Electrical Engineering,\\
University College London, WC1E 6BT, London, UK.\\
ryan.white.23@ucl.ac.uk}
\IEEEauthorblockA{\IEEEauthorrefmark{2}Defence Science and Technology Laboratory (Dstl),\\
SP4 0JQ, Salisbury, UK.}
}%

\maketitle

\begin{abstract}

This paper introduces "RadarML", a publicly available dataset of modulated radar pulses recorded over-the-air across multiple semi-independent receiver channels. The dataset was generated using the UCL ARESTOR platform which is a Radio Frequency System on a Chip (RFSoC) based Software Defined Radio (SDR). The dataset contains a library of around two million waveforms with varying centre frequencies, bandwidths, pulse widths and modulations, with 1,000 examples of each signal variation per receiver channel. The goal of this dataset is to enable comparative analysis of radar modulation classification techniques on real data and stimulate research into multi-channel signal detection and modulation classification methods.

\end{abstract}
\begin{IEEEkeywords}
Radar, ESM, ELINT, Software Defined Radio.
\end{IEEEkeywords}

\section{Introduction}

As utilisation of the Radio Frequency (RF) spectrum continues to grow, the density of emissions within any given frequency band increases. For Electronic Surveillance (ES) systems, the ability to reliably detect and classify radar signals has therefore become increasingly important. Modern radar systems, particularly software defined radars are waveform agile and capable of adapting parameters such as centre frequency, bandwidth, pulse duration, power and modulation on a pulse-to-pulse basis. This agility, combined with the proliferation of complex modulation schemes, creates a significantly more challenging electromagnetic environment for ES systems.

There are many publications that deal with the problem of signal detection and modulation classification of radar signals. However, many of these papers use different definitions and implementations of pulse modulation types making comparative analysis of techniques difficult, and very few utilise data that was captured experimentally. As far as the authors of this paper are aware, none use multiple co-located receiver channels simultaneously capturing the same radar pulses.

To address this gap in available data, we present "RadarML", a publicly available dataset of both frequency and phase modulated radar pulses that were captured experimentally over-the-air across multiple Analogue-To-Digital (ADC) channels using an RFSoC based system.

The aim of "RadarML" is to enable comparative analysis of radar modulation classification and parameter extraction techniques by serving as a standard reference dataset and to stimulate research into new radar electronic surveillance techniques that use multiple semi-independent receive channels to enhance detection and modulation classification performance.

\section{Dataset Description}\label{sec:dataset}

The dataset contains a broad range of representative radar waveforms with parameters as described below in Table \ref{tab:waveforms}. These waveforms vary in centre frequency, time duration or bandwidth (if applicable). Many of the waveforms are derived from open literature \cite{pace_2004}, however some more exotic examples such as non-linear chirps are included, which are more challenging than traditional radar waveforms for ES techniques. For each permutation of waveform there were 1,000 pulses transmitted.

\begin{figure*}[htbp]
\begin{center}
\includegraphics[scale=0.28]{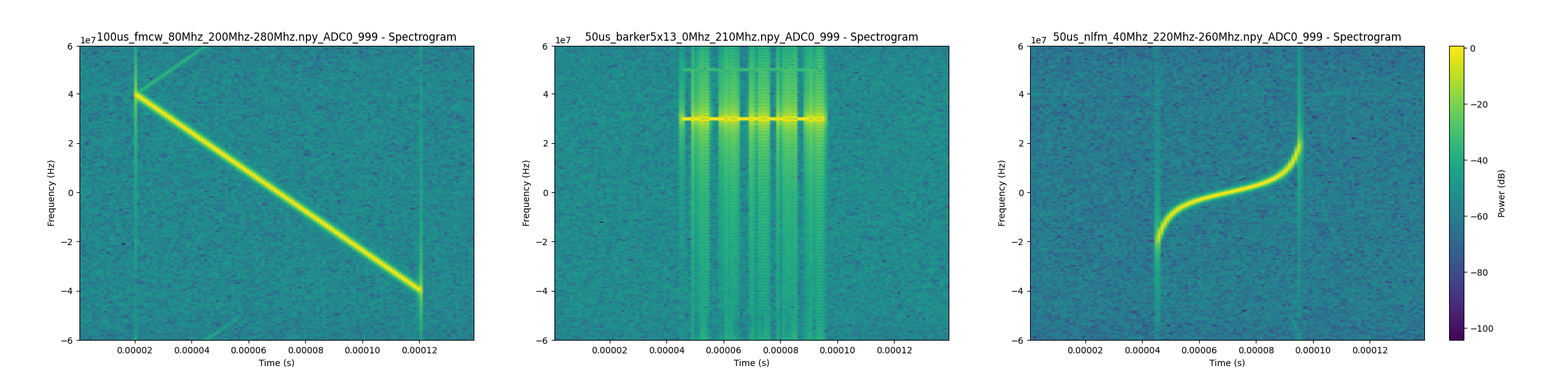}
\caption{Spectrograms generated from the raw data, showing examples of a linear chirp, Barker5x13, and non-linear chirp pulse modulations respectively.}
\label{fig1}
\end{center}
\end{figure*}

Some example spectrograms generated from the dataset are shown in Fig.~\ref{fig1}.

\begin{table}[h!]
    \centering
    \caption{Synthetic Waveform Library}
    \label{tab:waveforms}
    \setlength{\tabcolsep}{5pt}
    \begin{tabular}{|c|c|c|c|}
        \hline
        \textbf{Modulation} & \textbf{\makecell{Baseband\\Centre\\Frequencies\\(MHz)}} & \textbf{\makecell{Bandwidths\\(MHz)}} & \textbf{\makecell{Pulse\\Widths \\($\mu$s)}} \\ \hline
        CW             & -30, -10, 10, 30  & N/A              & 10, 50, 100   \\ \hline
        Linear FM      & -30, -10, 0, 10, 30     & +/- 10, 20, 40, 80   & 10, 50, 100   \\ \hline
        Exponential FM & -30, -10, 0, 10, 30             & +/- 10, 20, 40, 80   & 10, 50, 100   \\ \hline
        Quadratic FM   & -30, -10, 0, 10, 30             & +/- 10, 20, 40, 80   & 10, 50, 100   \\ \hline
        Tangential FM  & -30, -10, 0, 10, 30             & +/- 10, 20, 40, 80   & 10, 50, 100   \\ \hline
        4 CP-FSK       & -30, -10, 0, 10, 30  & 10, 20, 40, 80               & 10, 50, 100   \\ \hline
        Barker 5       & -30, -10, 10, 30  & N/A              & 10, 50, 100   \\ \hline
        Barker 7       & -30, -10, 10, 30  & N/A              & 10, 50, 100   \\ \hline
        Barker 11      & -30, -10, 10, 30  & N/A              & 10, 50, 100   \\ \hline
        Barker 13      & -30, -10, 10, 30  & N/A              & 10, 50, 100   \\ \hline
        Barker 5 $\otimes$ 11      & -30, -10, 10, 30 & N/A             & 10, 50, 100   \\ \hline
        Barker 5 $\otimes$ 13      & -30, -10, 10, 30 & N/A             & 10, 50, 100   \\ \hline
    \end{tabular}
\end{table}

For every waveform combination, the dataset contains a \texttt{NumPy} file with dimensions $[3, 1000, 33600]$, the dimension definitions are summarised in Table~\ref{tab:data_dims}.

\begin{table}[h]
\centering
\caption{Structure of each waveform dataset file}
\label{tab:data_dims}
\begin{tabular}{llp{5.7cm}}
\hline
Dimension & Size & Description \\
\hline
1 & 3 & ADC channels \\
2 & 1000 & Number of pulses acquired per waveform configuration \\
3 & 33600 & Interleaved I and Q samples per acquisition\\
\hline
\end{tabular}
\end{table}

The samples for each pulse are provided unprocessed, as digitised by the ADCs, in the \texttt{np.int16} format with no synthetic noise added. To evaluate ES algorithms across a range of signal to noise ratios (SNRs), additive white Gaussian noise (AWGN) can be added to the baseline samples at load time.

\begin{figure}[htbp]
\begin{center}
\includegraphics[scale=0.34]{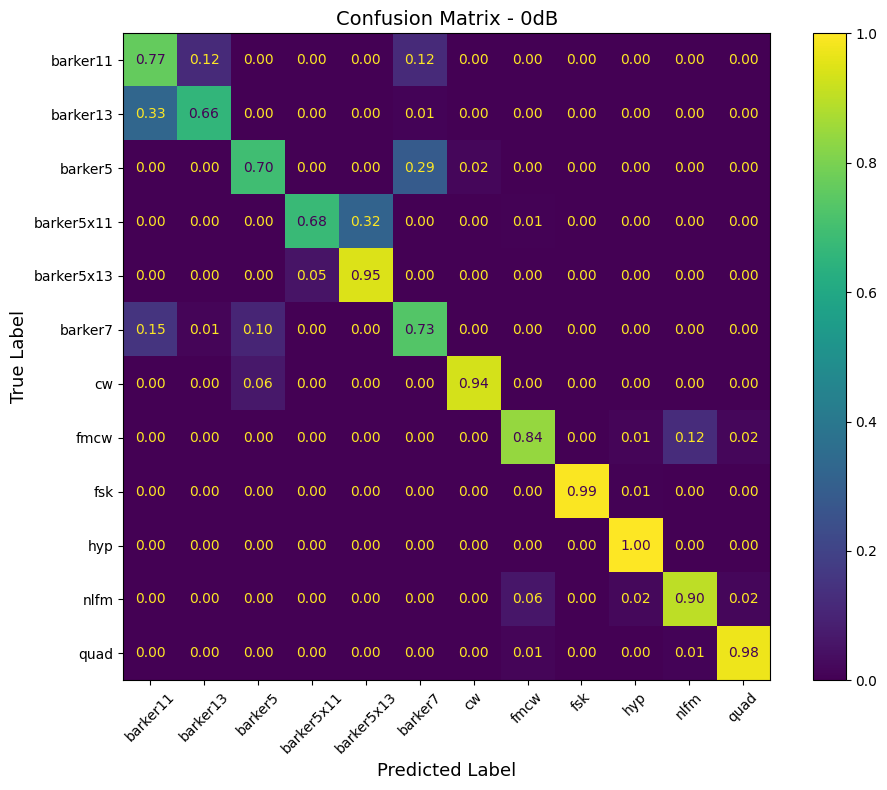}
\caption{Confusion matrix at 0 dB SNR for a VGG13 Model trained on ADC0.}
\label{fig2}
\end{center}
\end{figure}

Along with the data, a code repository is provided (see Section~\ref{sec:code}) which includes a \texttt{PyTorch} dataloader for viewing the raw data, and a wrapper dataloader that supports augmentations in SNR, frequency, phase and time to improve generalisation (as shown in Fig.~\ref{fig5}). A spectrogram transform and a baseline VGG13-style classifier are also included. Two \texttt{jupyter notebook}s demonstrate their use:

\begin{itemize}

  \item \textbf{dataloader\_demo.ipynb} - Provides a step-by-step demo of the dataloaders. It covers loading and viewing the raw pulses, comparing the same physical pulse across ADCs, balancing the number of pulses per class, applying each augmentation, and seeding for repeatable results.

  \item \textbf{vgg13\_training\_demo.ipynb} - Provides an end-to-end example that trains and evaluates the VGG13 classifier on spectrograms from a single ADC. Training uses pulses degraded from 30 dB to 0 dB SNR, with random frequency, phase and time shifts. The trained model is evaluated from 15 dB to $-$15 dB SNR, with per-SNR confusion matrices and macro F1 plotted against SNR.

  An example confusion matrix at 0 dB SNR generated from this script is shown in Fig.~\ref{fig2}.

\end{itemize}

\subsection{Waveform Definitions}

The equations that describe the instantaneous frequency of the frequency modulated chirps in the dataset are given below:

\begin{equation}
Linear~f(t) = f_0 + (f_1-f_0) * (\frac{t}{t_1})
\end{equation}

\begin{align}
Exponential~f(t) &= f_0 + (f_1 - f_0) \cdot \left( \frac{t}{t_1} \right)^{k} \\
\nonumber\text{where } k &= 3
\end{align}

\begin{equation}
Quadratic~f(t) = f_1 - (f_1-f_0) * \frac{(t_1/2-t)^2}{(t_1/2)^2}
\end{equation}

The derivation of the instantaneous frequency of the tangential chirp is given in \cite{nlfm} by Widyantara et al.

\begin{align}
Tan~f(t) &= \frac{f_0 + f_1}{2} + (f_1 - f_0) \cdot
\frac{\tan\left( \frac{2\beta\,(t - \frac{t_1}{2})}{t_1} \right)}{2\,\tan(\beta)} \\
\nonumber\text{where } \beta &= \arctan(\alpha), \quad \alpha = 4
\end{align}

In half of the frequency-modulated examples, the start $(f_0)$ and stop $(f_1)$ frequencies were swapped to generate the corresponding "down" chirps.

For the continuous-phase frequency-shift-keyed (CP-FSK) waveforms, four frequencies are chosen at random that satisfy the bandwidth and centre-frequency constraints provided. The pulse is split equally into sub-pulses which are each assigned one of the frequency values. To ensure spectral purity and minimise abrupt phase changes, the phase of each sub-pulse transitions continuously from the previous sub-pulse, maintaining phase continuity across frequency shifts.

In the binary Barker code implementation, the pulse is again split into sub-pulses, each of which is given a code value of "1" or "-1". Whenever the code changes between sub-pulses, the phase is shifted by 180 degrees. 

Some compound Barker codes are also present in the dataset. These are generated by taking the Kronecker product of two different Barker sequences to create a longer code, which is split into sub-pulses and phase shifted using the same method described in the previous paragraph.

\section{Hardware Description}\label{sec:hardware}

\begin{figure}[htbp]
\begin{center}
\includegraphics[scale=0.09]{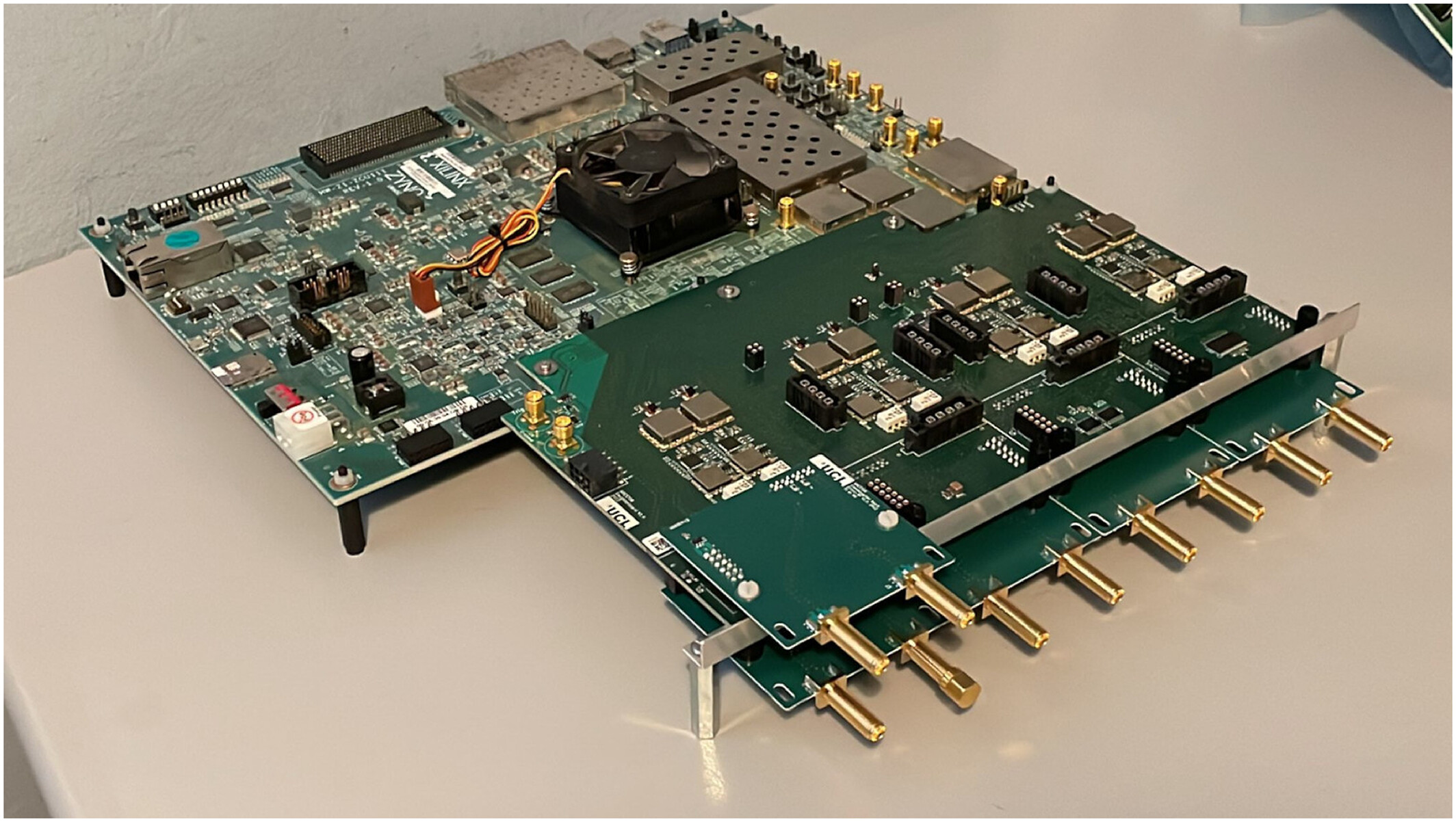}
\caption{The UCL ARESTOR System.}
\label{fig3}
\end{center}
\end{figure}

The dataset introduced in this paper was transmitted and received over-the-air in the 2.4 GHz ISM (Industrial, Scientific and Medical) band in an anechoic chamber using the UCL "ARESTOR" system shown in Fig.~\ref{fig3}, which is an SDR based on an RFSoC architecture. ARESTOR has enabled experiments in active and passive radar \cite{9905049}, Integrated Sensing and Communications (ISAC) \cite{isac1}\cite{isac2}\cite{isac3} as well as ES \cite{es1}\cite{es2}\cite{es3}. More details about the ARESTOR system and its hardware are provided in previous papers authored by the UCL Radar Group \cite{9784466} \cite{10149783}.

To generate the RadarML dataset, a single ARESTOR node acting as both the transmitter and receiver(s) was used to avoid the requirement of synchronising multiple nodes. A single DAC (Digital-to-Analogue) channel was used as the transmitter and 3 semi-independent ADC channels on different RF-Tiles were used for receive. Whilst the PLLs (Phase-Locked Loops) on each RF-Tile are locked to the same reference clock, no phase synchronisation (Multi-Tile Synchronisation) is performed between them, therefore the ADC channels are considered mesochronous.

The dataset waveforms were generated at an intermediate centre frequency of 240 MHz at a sample rate of 720 MS/s before interpolation by a factor of 8 by the DAC tile on the RFSoC to achieve a final sample rate of 5.76 GS/s. The waveforms are then digitally up-converted by mixing with a complex NCO (Numerically Controlled Oscillator) operating at a frequency of +2.2 GHz to enable transmission within the ISM band at a centre-frequency of 2.44 GHz.

The receivers on the RFSoC operate at a sample rate of 3.84 GS/s, which is then decimated by a factor of 32 in two stages ($8\times$ by the RF Tile and $4\times$  by further FPGA processing) to achieve a final sample rate of 120 MS/s. As the desired signal is sampled in the 2nd Nyquist zone, spectral inversion occurs. The actual intermediate frequency after sampling is given in \eqref{eq:if_second_nyquist}. The digitised signals are then digitally down converted using an NCO operating at -1.4 GHz, which mixes the centre-frequency of our target signals to baseband. 

\begin{equation}
f_\text{IF} = \lvert f_s - f_\text{RF} \rvert, \quad \frac{f_s}{2} < f_\text{RF} < f_s
\label{eq:if_second_nyquist}
\end{equation}

The receiver(s) are configured to record for 140 µs per acquisition, providing sufficient temporal coverage for all pulse durations in the dataset. Acquisitions are scheduled at 1 kHz to align with the transmitter's PRF of 1 kHz.

\begin{figure}[h]
\begin{center}
\includegraphics[scale=0.09]{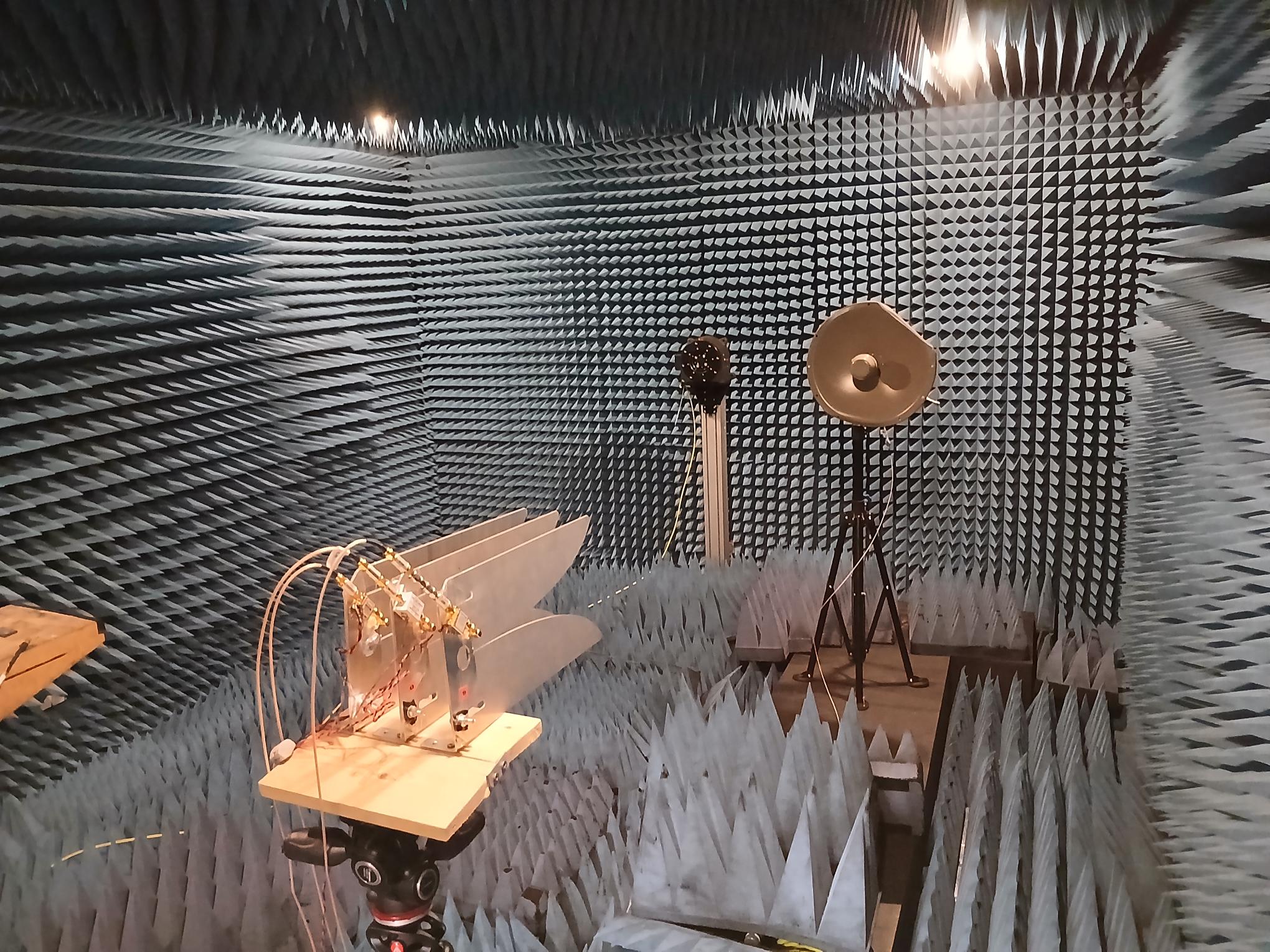}
\caption{Receiver setup in the anechoic chamber, showing the 3x co-located receive antennas and the transmit antenna.}
\label{fig4}
\end{center}
\end{figure}

To ensure each transmitted pulse is centred within its corresponding receiver acquisition window, a variable pre-transmission delay is applied based on the pulse duration:

\[
t_{\text{transmit delay}} = \frac{\left\lceil \left(1.4 \times 10^{-4} - t_{\text{pulse duration}}\right) \times 5 \times 10^5 \right\rceil}{10^6}
\]

\medskip

Each receive channel uses the same model of antenna (RFSPACE UWB3) and filter (Mini Circuits VBF-2435+), however, a different amplifier is used with each to introduce further variation into the digitised data. Attenuators are used to ensure each channel has roughly the same SNR regardless of the gain and noise factor of the amplifier chosen. All the information on the physical hardware attached to each DAC/ADC channel used in this setup are described below in Table \ref{tab:DAC_channels}.

The physical setup of the chamber is shown in Fig.~\ref{fig4}.

\begin{table}[h!]
    \centering
    \caption{DAC/ADC Channel Configurations}
    \label{tab:DAC_channels}
    \setlength{\tabcolsep}{5pt}
    \begin{tabular}{|c|c|c|c|c|}
        \hline
        \textbf{Port} & \textbf{Antenna} & \textbf{Amplifier} & \textbf{Filter} & \textbf{Attenuation} \\ \hline
        DAC0 & \makecell{L-COM\\HG2418DPD} & \makecell{Mini-Circuits\\ZX60-6013E-S+} & \makecell{Mini-Circuits\\VBF-2435+} & N/A \\ \hline
        ADC0 & \makecell{RFSPACE\\UWB3} & \makecell{Mini-Circuits\\ZX60-6013E-S+} & \makecell{Mini-Circuits\\VBF-2435+} & N/A \\ \hline
        ADC2 & \makecell{RFSPACE\\UWB3} & \makecell{Mini-Circuits\\ZJL-4HG+} & \makecell{Mini-Circuits\\VBF-2435+} & 1 dB \\ \hline
        ADC4 & \makecell{RFSPACE\\UWB3} & \makecell{Mini-Circuits\\ZX60-83LN12+} & \makecell{Mini-Circuits\\VBF-2435+} & 13 dB \\ \hline
    \end{tabular}
\end{table}

\section{Data and Code Availability}\label{sec:code}

\begin{figure*}[htbp]
\begin{center}
\includegraphics[scale=0.36]{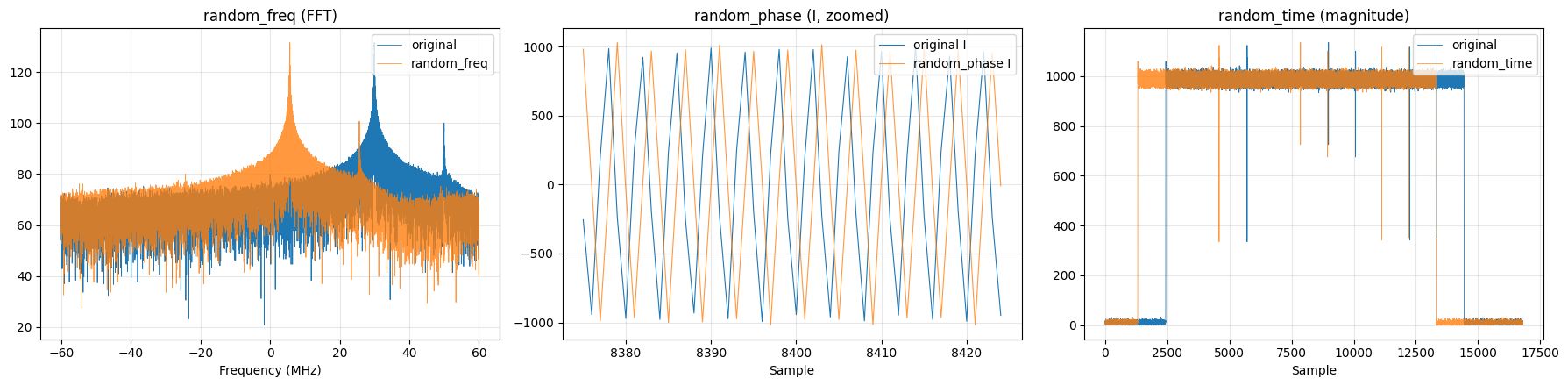}
\caption{A plot showing how frequency, phase and time augmentations affect the data by plotting the FFT and in-phase and magnitude component against time.}
\label{fig5}
\end{center}
\end{figure*}

The dataset and some source code to get started are publicly available, the links for which are provided below:

\begin{itemize}
    \item \textbf{Dataset:} The full dataset as described in this paper is available at 
    \url{https://doi.org/10.5522/04/30752767}. 

    \item \textbf{Example Processing Scripts:} The pre-processing and plotting scripts plus the example neural network as described in section \ref{sec:dataset} are available at \url{https://github.com/UCLRadarGroup/Radar_ML}. 

\end{itemize}

The dataset is released under the CC BY 4.0 license and the code is released under the GPL-3.0 license.

\section{Conclusions}\label{sec:Conclusions}


This paper has presented RadarML, a multi-channel dataset of representative radar waveforms captured over-the-air to support the evaluation of electronic surveillance and modulation classification techniques. The data was generated using the UCL ARESTOR platform in order to systematically produce large quantities of pulses over a broad range of modulations and pulse parameters. ARESTOR operated as both the radar as well as the ES receiver(s) in this case.

It is hoped that RadarML will enable reproducible comparison of algorithms and stimulate further research into multi-channel signal processing and fusion approaches.

\section*{Acknowledgments}\label{sec:acknowledgments}

This dataset was generated under an iCASE PhD Studentship funded by the Engineering and Physical Sciences Research Council (EPSRC) and the Defence Science and Technology Laboratory (Dstl). Prof. Ritchie’s time is supported by The Royal Academy of Engineering and Leonardo UK under the Research Chairs and Senior Research Fellowships programme.


\bibliography{EuRad24_rhwhite}

@book{pace_2004,
  author = {Pace, Phillip E},
  publisher = {Artech House},
  title = {Detecting and classifying low probability of intercept radar},
  year = {2004}
}

@INPROCEEDINGS{9784466,
  author={Peters, Nial and Horne, Colin and Ritchie, Matthew A.},
  booktitle={2021 18th European Radar Conference (EuRAD)}, 
  title={{ARESTOR}: A Multi-role {RF} Sensor based on the {Xilinx RFSoC}}, 
  year={2022},
  volume={},
  number={},
  pages={102-105},
  doi={10.23919/EuRAD50154.2022.9784466}
}

@INPROCEEDINGS{10149783,
  author={Peters, Nial J. and Horne, Colin P. and Amiri, Amin D. and Beasley, Piers and Ritchie, Matthew A.},
  booktitle={2023 IEEE Radar Conference (RadarConf23)}, 
  title={Modular Multi-Channel RFSoC System Expansion and Array Design}, 
  year={2023},
  volume={},
  number={},
  pages={1-4},
  doi={10.1109/RadarConf2351548.2023.10149783}
}

@INPROCEEDINGS{9905049,
  author={Ritchie, M. and Peters, N. and Horne, C.},
  booktitle={2022 23rd International Radar Symposium (IRS)}, 
  title={Joint Active Passive Sensing using a Radio Frequency System-on-a-Chip based sensor}, 
  year={2022},
  volume={},
  number={},
  pages={130-135},
  doi={10.23919/IRS54158.2022.9905049}
}

@ARTICLE{isac1,
  author={Temiz, Murat and Horne, Colin and Peters, Nial J. and Ritchie, Matthew A. and Masouros, Christos},
  journal={IEEE Transactions on Microwave Theory and Techniques}, 
  title={An Experimental Study of Radar-Centric Transmission for Integrated Sensing and Communications}, 
  year={2023},
  volume={71},
  number={7},
  pages={3203-3216},
  doi={10.1109/TMTT.2023.3234309}
}

@INPROCEEDINGS{isac2,
  author={Temiz, Murat and Peters, Nial J. and Horne, Colin and Ritchie, Matthew A. and Masouros, Christos},
  booktitle={2023 IEEE Radar Conference (RadarConf23)}, 
  title={Radar-Centric ISAC Through Index Modulation: Over-the-air Experimentation and Trade-offs}, 
  year={2023},
  volume={},
  number={},
  pages={1-6},
  doi={10.1109/RadarConf2351548.2023.10149620}}

@INPROCEEDINGS{isac3,
  author={Winter, Robert S. C. and Catherall, Aled and Masouros, Christos and Ritchie, Matthew},
  booktitle={2024 IEEE International Workshop on Technologies for Defense and Security (TechDefense)}, 
  title={Impact and Mitigation of Index Modulation on Radar Quality for a FMCW-Based Communication System}, 
  year={2024},
  volume={},
  number={},
  pages={485-489},
  doi={10.1109/TechDefense63521.2024.10863567}
}

@ARTICLE{es1,
  author={Horne, Colin and Peters, Nial J. and Ritchie, Matthew A.},
  journal={IEEE Access}, 
  title={Classification of LoRa Signals With Real-Time Validation Using the Xilinx Radio Frequency System-on-Chip}, 
  year={2023},
  volume={11},
  number={},
  pages={26211-26223},
  doi={10.1109/ACCESS.2023.3252170}
}

@INPROCEEDINGS{es2,
  author={White, Ryan and Horne, Colin and Ritchie, Matthew A.},
  booktitle={2024 21st European Radar Conference (EuRAD)}, 
  title={An Empirical Evaluation of PDW Extraction Techniques Against Weak LPI Pulsed Radar}, 
  year={2024},
  volume={},
  number={},
  pages={260-263},
  doi={10.23919/EuRAD61604.2024.10734975}
}

@INPROCEEDINGS{es3,
  author={White, Ryan and Winter, Robert S. C. and Horne, Colin and Ritchie, Matthew A.},
  booktitle={2025 IEEE International Radar Conference (RADAR)}, 
  title={Evaluation of YOLO for Automatic Radar Waveform Detection and Classification in Congested EME}, 
  year={2025},
  volume={},
  number={},
  pages={1-6},
  doi={10.1109/RADAR52380.2025.11031686}
}

@article{nlfm,
  title={Analysis of Non Linear Frequency Modulation (NLFM) Waveforms for Pulse Compression Radar},
  author={Muhamad Ridwan Widyantara and Sugihartono and Fiky Y. Suratman and Slamet Widodo and Pamungkas Daud},
  journal={Jurnal Elektronika dan Telekomunikasi},
  year={2018},
  url={https://api.semanticscholar.org/CorpusID:73544311}
}
\bibliographystyle{ieeetr}



\vspace{12pt}

\end{document}